\documentclass[12pt]{article}
\usepackage[french]{babel} 
\usepackage[T1]{fontenc} 
\usepackage[utf8]{inputenc}
\usepackage{algorithm}   
\usepackage{algorithmic}  
\usepackage[dvipsnames]{xcolor} 
\usepackage{slashbox}
\usepackage[normalem]{ulem} 

\usepackage{amsmath}
\usepackage{amsfonts}
\usepackage{subcaption} 
\usepackage{graphicx}
\usepackage[round]{natbib}

\begin{document}

	
	\begin{center}
		{\Large
			{Prediction model for rare events in longitudinal follow-up and resampling methods}
		}
		\bigskip
		
		Mathieu Berthe $^{1}$ \& Pierre Druilhet $^{2}$ \& St\'ephanie L\'eger $^{3}$ 
		\bigskip 
		
		{\it
			$^{1,2,3}$ Universit\'e Clermont Auvergne
			
			Laboratoire de Math\'ematiques Blaise Pascal UMR 6620 - CNRS
			
			Campus des C\'ezeaux
			
			3, Place Vasarely
			
			TSA 60026 \- CS 60026
			
			63178 Aubi\`ere Cedex
			
			$^{1}$ Mathieu.Berthe@math.univ-bpclermont.fr
			
			$^{2}$ Pierre.Druilhet@math.univ-bpclermont.fr
			
			$^{3}$ Stephanie.Leger@math.univ-bpclermont.fr
			
		}
	\end{center}
	\bigskip
	
	\begin{abstract}
		We consider the problem of model building for rare events prediction in  longitudinal follow-up studies. In this paper,  we  compare several resampling methods  to improve  standard regression models on a real life example. We  evaluate the effect of the  sampling rate on the predictive performances of the models.  To evaluate the predictive performance of a longitudinal model,  we consider   a validation technique   that takes into account time and corresponds to the actual use in real life.
		
	\end{abstract}
	

\section{Introduction}
Prediction models for rare events appears in many research fields  such as  economic \citep{eco},  politics  \citep{King}, fraud detection \citep{BoltonHand2002} or   bank regulation \citep{Calabrese}.
Modeling and predicting binary rare events  present several difficulties.  Strong imbalance between event and non-events induce  biased estimations and poor predictive performances, usually underestimating the probability of event  occurrences.  
In recent years, several strategies have been  proposed to improve misclassification. 
For example, \cite{King} propose an explanatory logistic regression model with bias correction in a case-control study.  
 \cite{Calabrese2} have developed a new     regression  model based on   extreme value theory.
More recently,  \cite{Haydemar} improve the learning function in SVM by       a low-cost post-processing strategy.

Another  way to the  improve predictive performance of a model with rare events is to  rebalance artificially  the dataset by  resampling methods. For example,   oversampling methods   creates artificially  new observations in the minority class, whereas undersampling methods  delete observations in the majority class. Hybrid methods  combine both oversampling and undersampling methods. 

The choice of  resampling rate, that is the final ratio between events and non-events, is a crucial point to improve predictive performance of the model. It is known that the optimal rate is highly dependent on the dataset \citep{compasampl2,Benchmarking}.  Futhermore, resampling methods induce additional randomness in the   dataset. The most common way to reduce this extra-variablity is to use aggregation methods  \citep{Breiman1996BaggingP}.
Other strategies to improve classifiers with rare events have been considered, such as  weighting training instances \citep{Pazzani1994ReducingMC} or using different misclassification costs for minority and majority events  \citep{gordon}.

The aim of the paper is to compare several resampling and aggregation methods on a real-life longitudinal follow-up study. We discuss the way to evaluate predictive performance in the case of longitudinal studies and then choose the optimal sampling rate adapted to our data set.

In Section \ref{section.logregrare}, we review  resampling  and ensemble based methods.  We also discuss the way to evaluate the  predictive performance adapted to longitudinal follow-up.
In Section \ref{section.realexample}, we  compare  several strategies applied  to    a real life  example: we have followed a soccer teams during one year and we aimed to evaluate the risk of muscle injury before each match.    We  discuss the crucial choice of  the sampling rate and the effect of aggregation methods.   We also show that SMOTE methods \citep{SMOTE2002} applied to our dataset performs poorly.

\section{Prediction models and sampling methods}

\label{section.logregrare}

In this section, we present  several resampling methods  combined with aggregation to improve the predictive performance  of a logistic regression.

\subsection{Standard logistic regression}

Here, we recall the bases  of logistic regression. For an individual $i$ $(i=1,...,n)$, let  $x_i$ be the $k+1$-vector of the  $k$ explanatory variables plus the constant and  let $y_i \in \lbrace 0,1\rbrace$ be  the binary response  which follows a Bernoulli distribution  with   parameter $\pi_i=P(y_i=1 \mid x_i)$. In the standard logistic regression, it is assumed that

\begin{equation} \label{pi}
	\pi_{i}=\frac{1}{1+e^{-x_{i}' \beta}}
\end{equation}
where $x'$ is the transpose of $x$ and $\beta'=(\beta_0,\beta_1,...,\beta_{k})$ is the  vector of   unknown parameters,    usually  estimated by maximum likelihood (see e.g.  \cite{GLM}). The asymptotic variance of $\widehat{\beta}$ is  

\begin{equation}V(\hat{\boldsymbol{\beta}})=\left[\sum_{i=1}^{n} \pi_{i}\left(1-\pi_{i}\right) \mathbf{x}_{i} \mathbf{x}_{i}  ^{\prime}\right]^{-1}
\end{equation}

For a new individual $x$, the probability of the event $y=1$ is predicted by
$$
\widehat\pi(x)=\frac{1}{1+e^{-x'_i\widehat\beta}}=\widehat{ \mathbb P}(y=1 \mid x)
$$

~

When the  dataset contains few events, say less than 5\%, it is known that logistic regression  underestimate the probability of  events and then    poor predictive  performances (see \citep{King})

\subsection{Balancing unbalanced dataset}
\label{resampling}
To overcome   drawbacks induced by the unbalanced datasets, several sampling methods can be used to artificially  rebalance  the  dataset. 
Several   resampling methods on real data are compared in  \cite{compasampl1,compasampl2,compasampl3, Benchmarking}.    \cite{DrummondandHolte2003} show that  oversampling is better than undersampling and   \cite{Japkowicz}   that random oversampling or undersampling methods improve substantially the predictive  performance of the models so  that    more sophisticated oversampling or down-sizing methods   approaches appear  unnecessary. All these studies show  that the best resampling method is highly dependent on the dataset.

In this section, we review   the most common sampling methods, which can be used alone or combined.  
\subsubsection{Undersampling methods}

The first way to rebalance an unbalanced dataset is to  reduce  the number of observations in the majority class (non-events). A random  undersampling with rate   $r$,  $0<r<1$,  creates a new dataset by removing at random   from the initial dataset  a proportion $r$ of observations from the majority class. If $r=0$, then all the observations of the majority class are kept. If $r=0.7$, then $70\%$ of the observations of the majority class are removed.

In the case of very rare events,  \cite{King} propose to used case-control designs \citep[see also][]{Breslow}. This  strategy is equivalent to  selecting randomly one  non-event  for every event, resulting in  a completely balanced dataset. In that case, if the proportion of events is $p$, then the rate of the undersampling is  $r=(1-2p)/(1-p)$. Another more sophisticated  strategy has been proposed in \cite{Tomeks1}: for each event, the idea is to remove  a non-event   that form a Tomek link.   \cite{compatomek}  considers situations where Tomek link methods does not   guarantee a performance gain.

The main drawback of undersampling methods is the loss of information when the number of removed observations is large. In Section \ref{subsecton.agreg}, we consider aggregated methods that limit this loss of information.

\subsubsection{Oversampling methods}

At the opposite of undersampling methods, oversampling methods increase artificially  the number of observations in the minority class (events). A random oversampling with rate $(a\!:\!b)$ creates  new observations  by   duplicating at random  observations in the minority class   until there are  $a$ non-events for $b$ events in the new dataset. An oversampling $(1\!:\!1)$ results in a completely balanced dataset. An oversampling $(2\!:\!1)$ results in a dataset with 2 non-events for 1 event.

\begin{figure}[h!]	
	\centering
	\includegraphics[scale=0.6]{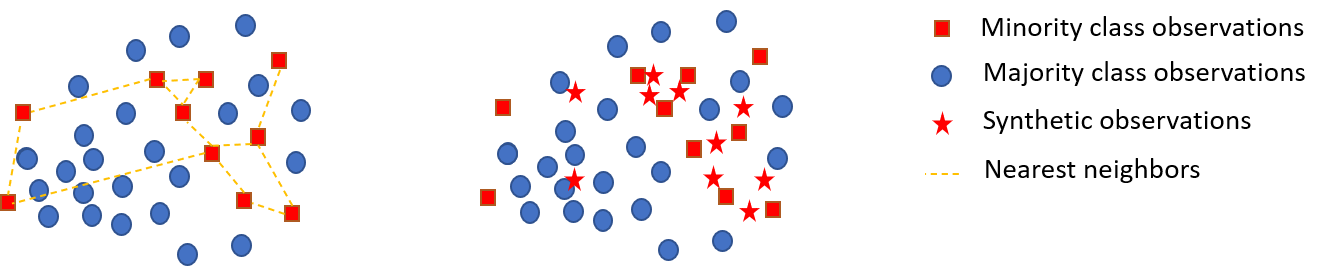}
	\caption{ SMOTE: for each event, one synthetic observation is created from one of the two nearest neighbors at random.}
	\label{fig.smote}
\end{figure}

SMOTE \citep{SMOTE2002} is a more sophisticated method that creates   synthetic observations in the minority class as follows: for each event observation, choose at random one of  the $k$   nearest neighbors that belongs to the minority class, with $k$  fixed.  The new synthetic  observation is chosen at random between these  two observations. It is also possible to reiterate the process to increase the oversampling rate. Figure \ref{fig.smote} shows the effect of SMOTE with $k=2$ and with one synthetic observation  generated by events.

\subsubsection{Hybrid sampling}

It is known  that these methods have some cons. Random undersampling can discard potentially useful data, whereas random oversampling creates exact copies of existing instances that may induce overfitting. 
To overcome these features, a solution is to mix undersampling and oversampling methods.  For example, a random undersampling method with rate $c$ combined  with a $(a\!:\!b)$-oversampling method consists in removing at random a proportion $c$ of non-event and then perform an oversampling  to obtain $a$ non-events for $b$ events.

As a remark,   random over/under sampling methods can be seen as weighted logistic regressions \citep{Manski}   where the weights are random.  For  the resampled dataset,  the log-likelihood of the logistic regression can be written:

\begin{align*}
	\ln L_{w}(\boldsymbol{\beta} | \mathbf{y})  
	& =  \sum_{i\,/\,y_{i}=1} w_i \ln \left(\pi_{i}\right)  + \sum_{i\,/\, y_i=0} w_i  \ln \left(1-\pi_{i}\right)
\end{align*}
where the weight $w_i$ is the number of replication of $x_i$ for $y_i=1$ in the random oversampling process  and $w_i=0/1$ for $y_i=0$ in the random  undersampling process. 

\subsection{Ensemble-based methods}
\label{subsecton.agreg}
Each sampling method  described above induces a supplementary part of randomness in the dataset and therefore more variability in the predictions. Ensemble-based methods are the  most common way to reduce this variability. The idea is to create $K$    datasets from the same resampling scheme and to aggregate the  predictors. Therefore, for a new individual $i$ with covariate $x_i$, the predicted probability of event $\widehat{\pi}_i$ is given by 
$$
\widehat \pi_i =\frac{1}{K}\sum_{k=1}^K\widehat{\pi}_i^{[k]} ,
$$
where  $\widehat{\pi}_i^{[k]}$ is the predictor obtained from the k$^{th}$ dataset. The choice of $K$ will be discussed in Section \ref{aggregationreallife}.

As a variant, when using a pure oversampling methods, the  non-events may be replaced by $K$ bootstrap samples, similarly to  Bagging \citep{Breiman1996BaggingP}. In the same way, when using  a pure oversampling   method, the events may be replaced by  a bootstrap sample of them. The effects of this bootstrap variant on the aggregated predictors are displayed in Table \ref{fig.Ensemble}.

\subsection{Predictive performance evaluation in longitudinal follow-up}
\label{subsection.evaluationmetrics}
\begin{figure}
	\begin{center}
		\includegraphics[scale=0.55]{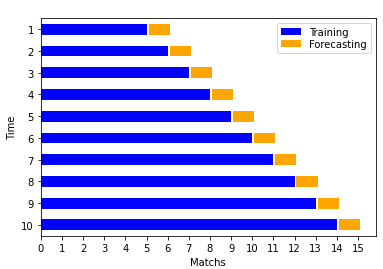}
	\end{center}
	\caption{Longitudinal validation.}
	\label{fig.longival}
\end{figure}

To evaluate the predictive performance of a model,   training and test datasets should be chosen carefully.
In  longitudinal follow-up studies, events are highly dependent on the past and change the future.   In this context, it is impossible to use standard  validation strategies like cross validation or random split of  the dataset   into learning and  test datasets. Indeed, with this strategy, the risk is to confuse  causes and consequences and to overestimate predictive performances. Therefore, it is more natural to use   a longitudinal strategy (see Fig. \ref{fig.longival}) that corresponds to the way the models are used in real life: at   time $t$, we only use  previous information    to predict the risk $\widehat \pi_{ti}$ to have an event on the individual $i$, then we compare our prediction with the real observation  $y_{ti}$. At the end, we have a  collection of $(\hat\pi_{tj},y_{tj})$, $t=1,...,T$ and $i=1,...,I$.

The usual way to compare  the ability of several  models  to predict a binary response is to compare their  ROC curves, AUCs or Peirce indices.  We recall that a ROC curve  is a parametric curve defined as follows:  for a given   threshold  $0\leq \gamma\leq 1$, we  predict $y_{ti}$ by    $\hat y_{ti}=0$ if $\widehat \pi_{ti} <\gamma$ and $\hat y_{ti}=1$ if $\widehat \pi_{ti}\geq \gamma$. Then, we compare the predicted response $\hat y_{ti}$ with the real outcome $y_{ti}$.  The sensitivity and the specificity, which depend on $\gamma$  are defined by  
\begin{equation*}
	\textrm{sensitivity}(\gamma)=\dfrac{TP}{TP+FN}
	\quad\quad \textrm{specificity}(\gamma)=\dfrac{TN}{TN+FN}\end{equation*}
with $TN$, $TP$, $FN$, $FP$ are the number of true negative, true positive, false negative, false positive. For example $TP=\#\{y_{ti}=1 \;,\;\hat y_{ti}=1\} $. The ROC curve is therefore the parametric curve $\{(\mathop{\textrm{sensitivity}}(\gamma), 1-\mathop{\textrm{specificity}}(\gamma))\;;\; 0\leq\gamma\leq 1\}$. As shown in  \cite{Raeder}, the choice of an evaluation metric plays an important role in   learning on unbalanced data. From the ROC curve,  we can derived two global metrics: the area under  ROC curve (AUC)    and  the Pierce index (PI)  defined by   
\begin{equation*}
	\textrm{PI}=\max_{\gamma\in [0,1]}\{\textrm{sensitivity}(\gamma) + \textrm{specificity}(\gamma) -1 \}.
\end{equation*}
which is particularly adapted to rare events.

The Pearce index   represents a good compromise between sensitivity and specificity. It can be shown  that $\textrm{PI}=1-d^*$, where $d^*$ is the  Manhattan distance between the point (0,1) and its closest point  on the ROC curve. It is also the euclidean distance between  the further point on the ROC curve from the diagonal, up to a factor $\sqrt 2$. 
The  model with the  highest AUC or PI will be considered as the best predictive model.

\section{Comparison of resampling methods in a  real life longitudinal follow-up.} 
\label{section.realexample}	
In this section, we apply and compare the methods described in Section \ref{section.logregrare} in a real life situation. We have followed a soccer teams of the french Ligue 1 Championship during the season 2018-2019.  We aim to build a  model that  evaluate  the individual risk of non-contact muscle injury for each player before each match. To build the model, we use the data collected during the seasons 2015-2018 and the season 2018 until the match. A review of football player injury prediction methods can be found in \cite{Ley2021}. Several predictive methods are compared in  \cite{CareyOWCCM2018}. 
From \cite{epi2}, the average  incidence of  muscle injuries for a player during a match is about 4\%. In our dataset we observe a similar rate, so that non-contact muscle injuries are considered as  rare events.

~

To evaluate the predictive performances of the model, we use the longitudinal validation described in Section \ref{subsection.evaluationmetrics}. Before each match, we predict the risk of muscle injury for each player $i$ based on all preceding observations. Then, we   and compare the prediction with the real outcomes,  that is, muscle injury or not of player $i$ during the match.   Of course, players that do not play the  match are not considered. 

~

During the season, 50 matches had been played and 16 non-contact muscle injuries have been observed.
To train the model before the first match, we use the data collected during the seasons 2015-2018. Then, iteratively,  we use   the data collected  until the day before each match of the season 2018-2019 to predict the probability of injury for the next match.

\subsection{The dataset}
\label{section.dataset}
The dataset include 42 soccer players on which data are collected daily and during matches. After  each match,  the response variable  is observed : $y=1$ if an injury is observed and  $y=0$ otherwise.  For each player, we have the following  covariates that are considered in the literature as risk factors.
\begin{itemize}
	\item[-] Cumulative workload  during training and matches over 21 days. 
	\item[-] Cumulative playing time over 21 days.
	\item[-] Recovery time: number of days since the last match.
	\item[-] Risk of relapse:  ratio between the number of days
	disability due to injury and the average number of days
	of disability in the team. It aims to quantify the risk of
	relapse after an injury.
	\item[-] Acceleration ratio: ratio between the number of accelerations
	performed over the 7 days preceding the match and the number of accelerations performed
	on the 21 days preceding the match
	\item[-] Deceleration ratio: ratio between the number of deceleration
	performed over the 7 days preceding the match and the number of deceleration performed
	on the 21 days preceding the match
	\item[-] Speed ratio: ratio between the average speed over the 7 days preceding the
	match and the average speed over the 21 days preceding the match.
	\item[-] Player ID:  player identifier.
\end{itemize}
Workload, Cumulative playing time  and  Recovery time allow to quantify player activity. Acceleration, deceleration and speed ratio are used to assess the player sport performance
before the match. 
Another important covariate is the player ID. In usual longitudinal studies,  the aim   is to extrapolate the model  on other individuals. Therefore  individuals (here,  players)  are considered as  random effects. In our case, we want to predict future observations on the individuals  that are included in the studies. Therefore,  players are considered   as fixed effect, allowing to personalize    the risk of injury.   We will not consider interaction between factors, since they have not shown,  in preliminary studies, any improvement of the predictive ability of the  models, mainly due to overfitting.

\subsection{Comparison of resampling methods}
In this section, we compare the predicitive performance of  several resampling strategies applied to logistic regression. The performances metrics are evaluated on the 50 matches played during  the season 2018-2019, by using  the longitudinal validation described in Section \ref{subsection.evaluationmetrics}. Several resampling methods are evaluated: undersampling alone, undersampling + bootstrap on events,  oversampling alone, oversampling + bootstrap on the events, both oversampling and undersampling.
When several sampling strategies  are combined, we first use  undersampling, then oversampling or SMOTE.

\subsubsection{Effect of  sampling rates on   predictive performances}
\label{subsec.samplingrate}
\begin{figure}
	\centering
	%
	\includegraphics[width=1\textwidth]{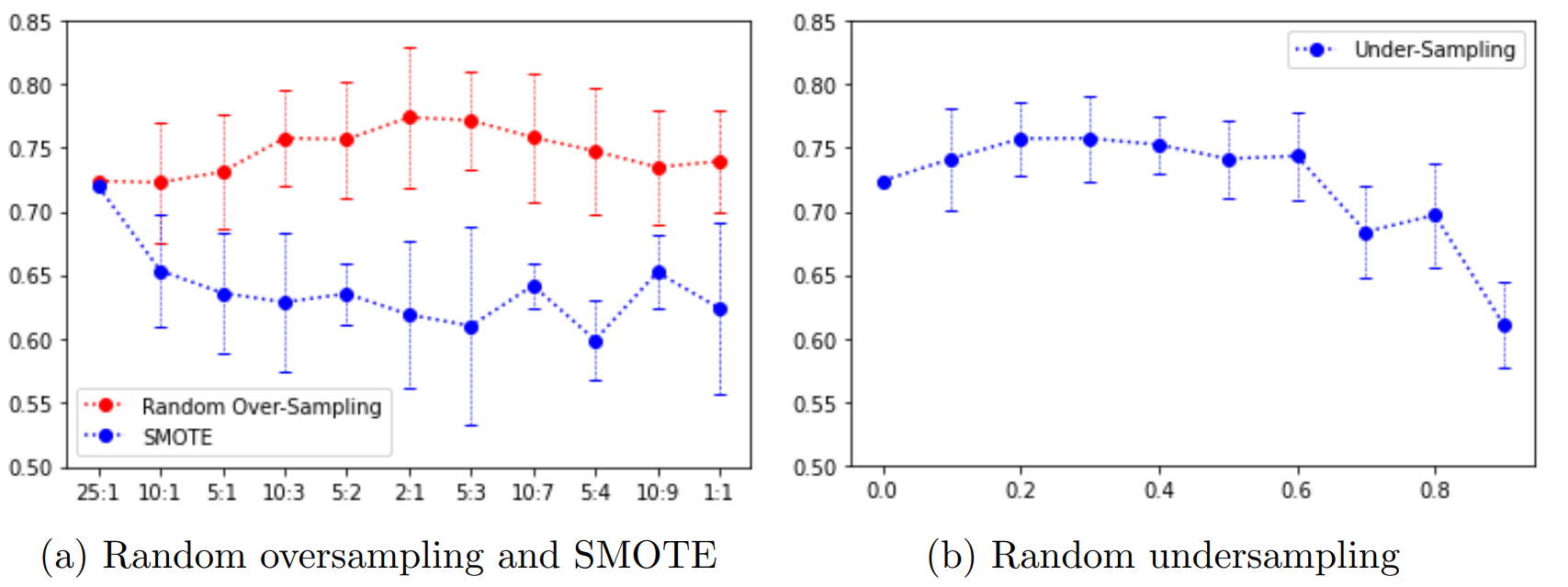}
	\caption{Average AUC over 15 runs with twice  the  standard deviation    against  sampling rates.}
	\label{fig.samplingrateAUC}
\end{figure}

Here, we  evaluate the effect of the balancing rate on   AUC in  random oversampling, SMOTE and  random  undersampling methods applied to logistic regression.  The results are displayed in Fig. \ref{fig.samplingrateAUC}. Each method is  run 15 times. Then, we compute the average   AUC over the runs. We also compute  the standard deviation for both metric.  
Note that the initial dataset imbalance  is $(25\!:\!1)$, that is 25 non-events for 1 event.


For random oversampling (Fig. \ref{fig.samplingrateAUC}.a, red line), the average  AUC increases   from $0.72$ to $0.77$ when the sampling rate goes from $(25\!:\!1)$  to   or $(5\!:\!3)$. Then,  the AUC decreases, probably due to an overfit on the events.    For  SMOTE (Fig. \ref{fig.samplingrateAUC}.a, blue line) the effect  on    AUC is always negative. This is mainly due to events that are isolated in the covariate  space and therefore create synthetic events in the middle of non-events: for example,  in    Fig. \ref{fig.smote}   two isolated events on the left induce two synthetic events  in the middle of a cluster of non-events.  

In Fig.  \ref{fig.samplingrateAUC}.b,  we can see that undersampling methods  slightly improve the average AUC    for an undersampling rate between 0.2 and 0.3 with a AUC gain   about 0.03. When the sampling rate is too large, say   greater than 0.7 for our dataset, the  predictive performance of the model  worsen since too many non-event individuals are removed.

\subsubsection{Comparison of several pure and hybrid resampling methods}\label{subsec.sampling}

\begin{table}[h!]
	\centering
	\includegraphics[scale=0.6]{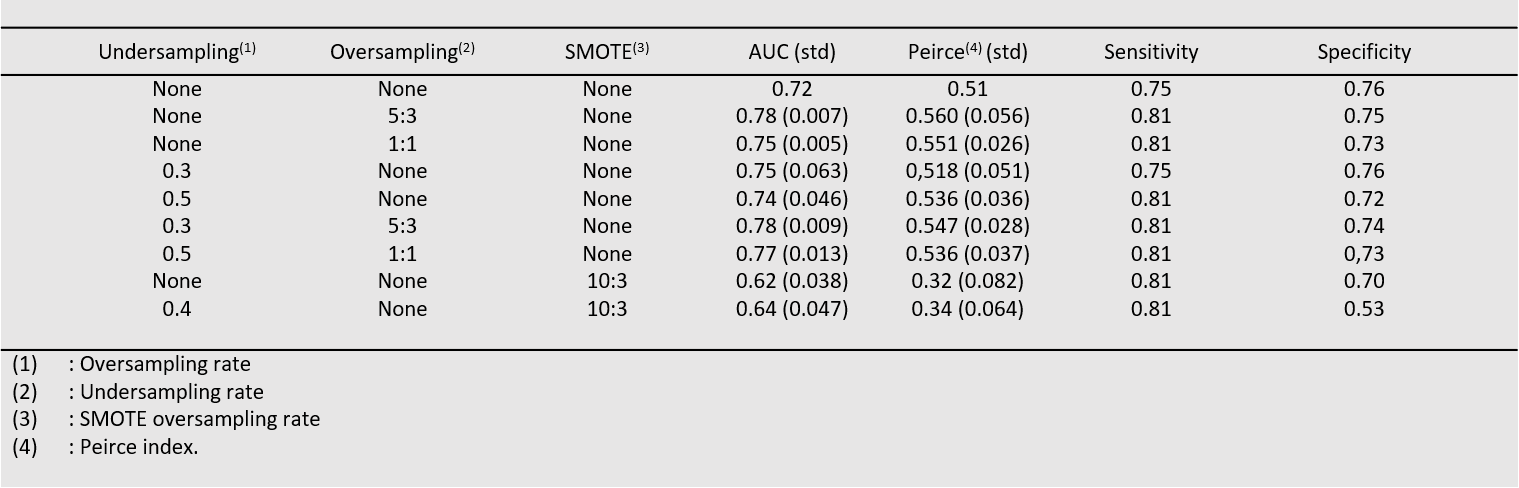} 
	\caption{Comparison of average AUC and Peirce index with standard deviation (std) for several resampling methods.}
	\label{fig.ressampling}
\end{table}

\begin{figure}[h!]
	\centering
	\includegraphics[scale=0.65]{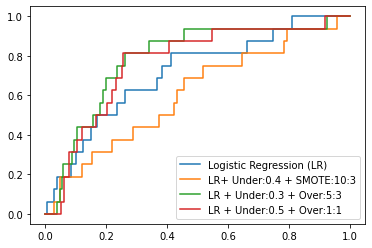}
	\caption{Comparison of ROC curves for several  resampling  methods with aggregation.}
	\label{rocurve}
\end{figure}

Here, we compare  random oversampling and undersampling methods studied in Section \ref{subsec.samplingrate} with hybrid methods,  SMOTE or plain logistic regression. Again, for each strategy, we run 15
times the model. So, we obtain an average AUC and  Peirce index with related  standard deviations. Sensitivities and specificities   are calculated for the  run whose Peirce index  is  the closest   to the average.
The results are   displayed in Table \ref{fig.ressampling} and for some models,  ROC curves are   displayed in Fig. \ref{rocurve}. 

The plain  logistic regression, i.e.  without additional resampling  method,  has  an AUC equal to  0.72 and a  Peirce index equal to 0.510 with a sensitivity   $0.75$ and a specificity $0.76$. Random oversampling improves the   prediction performance for a large range of  sampling rates.  For example, an oversampling    rate of $(5\!:\!3)$   gives average AUC and Peirce  equal to     $0.78$  and $0.56$, the sensibility increases to  $0.75$ whereas the specificity slightly decreases from to $0.75$. 
As already seen in Section \ref{subsec.samplingrate}, SMOTE methods give poor results and   undersampling should be used with caution, only with a  small removal rate.

In conclusion, for our dataset, the  resampling methods with highest  AUC and Peirce index are pure  random    oversampling $(5\!:\!3)$ followed by   hybrid undersampling $0.3$ /  oversampling $(5\!:\!3)$. Note that the second method  has  a slightly lower average Peirce index (0.547) for the same average AUC.

\subsubsection{Ensemble-based methods}
\label{aggregationreallife}
Resampling methods add randomness in the output. Ensemble based methods,   described in section \ref{subsecton.agreg}, aim to stabilize the model and in some situations to improve the predictive performance, similarly to Bagging methods. 
There is no consensus about the right number of   aggregations, which is  usually between 20 and 100 for Bagging methods \citep{Breiman1996BaggingP,Buhlmann}, depending on the dataset.

\begin{figure}[h!]
	\centering
	
	\includegraphics[scale=0.6]{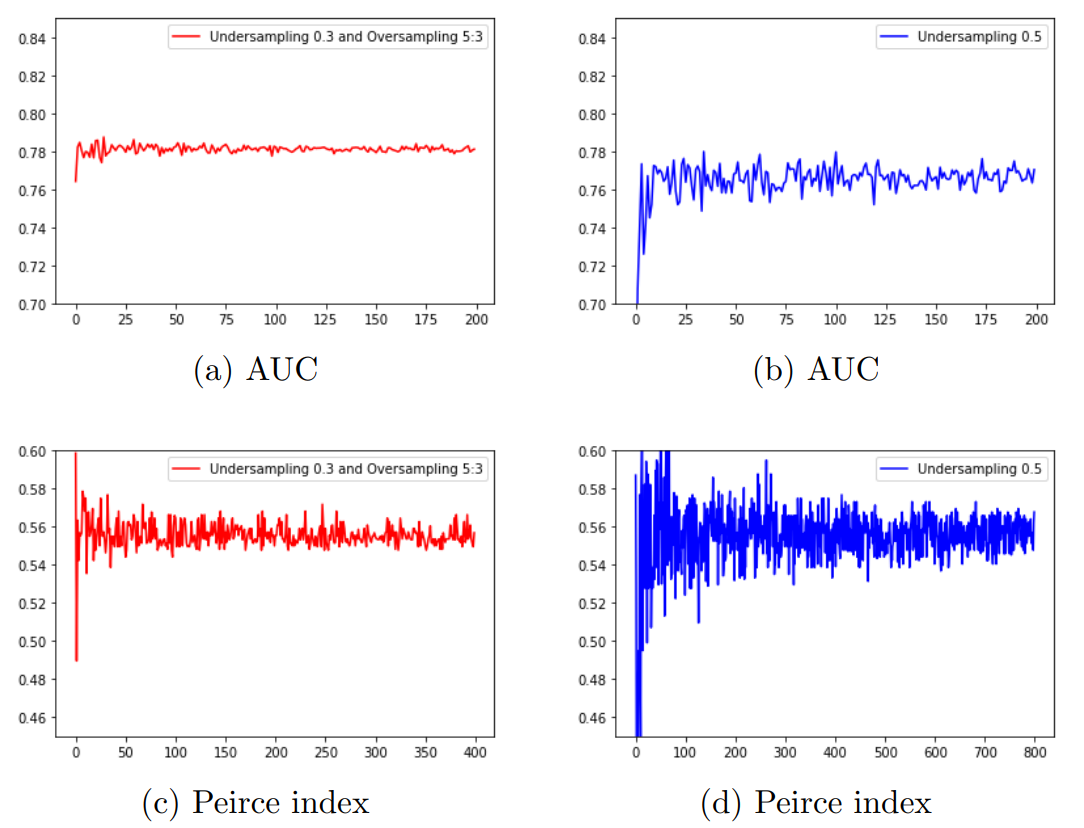}
	\caption[ AUC variation ]
	{\small AUC and Peirce index variation against the number of aggregation in ensemble-based methods for two resampling methods.} 
	
	\label{fig.aucbagging}
\end{figure}

To evaluate the  effect of the number of  aggregations needed to stabilize the prediction for our dataset, we display, in Fig. \ref{fig.aucbagging},   AUC and Peirce index against  to the number of aggregations for  two of the  best models obtained in Section \ref{subsec.sampling}: the first one is an hybrid undersampling with $r= 0.3$ and overampling $(5\!:\!3)$ and the second one is an undersampling with $r=0.5$ combined with a bootstrap  sampling on the events. 
For the two models,   AUC is stabilized after 20 iterations (Fig.  \ref{fig.aucbagging}.a and \ref{fig.aucbagging}.b) whereas  Peirce index needs more iterations to be stabilized (Fig.  \ref{fig.aucbagging}.c and \ref{fig.aucbagging}.d) . 

~

To save computer time, we now compare ensemble-based methods with   20 iterations for the  models used in Section \ref{subsec.sampling}. The results are displayed in Table \ref{fig.Ensemble}, line 1-6,   whose means and standard deviations of AUC and  Peirce index, sensitivity and specificity are obtained in the same way as in Table \ref{fig.ressampling}. We omit  SMOTE methods that have shown poor results.

It can be observed  that the main effect of aggregation methods is to reduce the variability of AUC and Peirce index. For example,   for  random undersampling with rate $0.3$, the standard deviation of AUC decreases from 0.063 to $0.005$. For   random oversampling $(5\!:\!3)$ it decreases from $0.007$ to $0.002$.
The effects of aggregation methods on the mean AUC and Peirce index depend on the resampling methods.
For undersampling,     aggregation methods improve slightly the average AUC and Peirce index, whereas there is no significant effect for oversampling. 

In table \ref{fig.Ensemble}, line 8-11, we have considered a bootstrap of the events when an undersampling method is used and a bootstrap sample of the non-events when an oversampling methods is used. It is seen that   Bootstrap has no significant  effect on the mean AUC and Peirce index, but increases their variability. In line 7 of the same table, we have performed a stratified bootstrap on the events and non-events. The predictive performance is better than that of  the plain logistic regression but lower than over/under sampling or hybrid methods with optimized rate.

Among all the models considered here, the   best predictive models are  the hybrid models    undersampling 0.5 and oversampling $(1\!:\!1)$ or undersampling 0.3 and oversampling $(5\!:\!3)$.

\begin{table}[h!]
	\centering
	\includegraphics[scale=0.7]{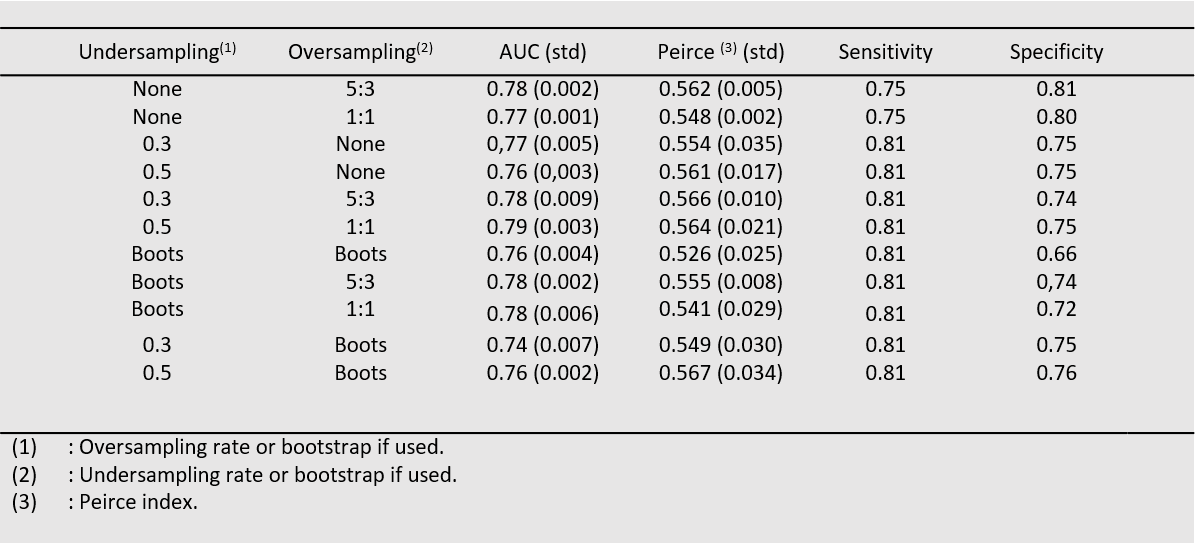}
	\caption{Effect of the ensemble-based methods on the mean and standard deviation of AUC and Peirce index for several resampling methods.}
	\label{fig.Ensemble}
\end{table}

\subsubsection{Longitudinal validation vs cross-validation}

Longitudinal validation described in Section \ref{subsection.evaluationmetrics} corresponds to the way the model is used in practice. It is therefore the most relevant method to evaluate the predictive performance.

Usual cross-validation methods such as  leave-one-out cross-validation (LOOCV) use future information to predict the outcome. For example, for our data set analyzed with  plain logistic regression,  i.e. without resampling methods,    AUC and Peirce indices obtained by LOOCV    are equal to  0.84 and 0.665 whereas  they are equal to  0.72 and 0.51 for  longitudinal validation. For  the best strategy found in 3.3.2, i.e. 0.5 undersampling  followed by  oversampling (5:3) with aggregation,    AUC and Peirce index are equal to 0.85 and   0.672 for LOOCV and to 0.78 and  0.566 for longitudinal validation. So, we can see that  LOOCV overestimate the true predictive performance of the models.

Another validation strategy consists in using the dataset based on the seasons 2015-2018 to train the model and  the dataset of the season 2018-2019 to test the model  \cite[see, e.g.,][]{CareyOWCCM2018}. This approach is relevant if it is not possible to update the model with fresh data or if we want to use the model for other individuals or players. However, in the case of   individual follow-up, the model  loose  information from the near past. For example,  with this validation approach, AUC and Peirce index are equal to 0.650 and 0.25 for the plain logistic regression 0.681 and 0.31 for the hybrid undersampling 0.5, oversampling $(5\!:\!3)$.
We can see  that this validation strategy tends to underestimate the predictive performance of the model as it is used in practice.

\section{Conclusion}

We have shown   how resampling methods can improve subtantially  predictive models for rare events. 
The best  resampling method  and the optimal sampling rate are specific to each dataset. Most often they are  calibrated by cross validation. However, in the case of longitudinal follow-up, usual cross-validation methods tend to overestimate the predictive quality of the model. Therefore,  it is important to use  a  validation method adapted to longitudinal follow-up. 

Pure random oversampling  or hybrid under/oversampling with optimized  sampling rate appear to be the most effective method to improve a logistic regression for rare events. 
SMOTE was ineffective for our dataset structure, mainly due  to  isolated events in the space of explanatory variables. 
Moreover, ensemble-based methods and predictor aggregation reduce the effects of the variability  of resampling methods onto the   predictors.

\subsection*{Acknowledgment} 	
The authors are grateful to     Olivier Brachet (Innovation Performance Analytics)    for having  provided the dataset.

\subsection*{Funding}

This research has been supported by the European Regional Development Fund and the Region Auvergne-Rhone-Alpes
\bibliographystyle{apalike} 
  \bibliography{biblio_methodo}
\end{document}